# Antiferromagnetic Skyrmion based Energy-Efficient Leaky Integrate and Fire Neuron Device

[1]Namita Bindal, [1]Ravish Kumar Raj, [2]Md Mahadi Rajib, [2,3]Jayasimha Atulasimha, and [1]Brajesh Kumar Kaushik

[1]*Department of Electronics and Communication Engineering,*
*Indian Institute of Technology Roorkee, Roorkee, Uttarakhand, India 247667*

[2]*Department of Mechanical and Nuclear Engineering,*
*Virginia Commonwealth University, Richmond, VA 23284, USA*

[3]*Department of Electrical and Computer Engineering,*
*Virginia Commonwealth University, Richmond, VA 23284, USA*

The development of energy-efficient neuromorphic hardware using spintronic devices based on antiferromagnetic (AFM) skyrmion motion on nanotracks has gained considerable interest. Owing to its properties such as robustness against external magnetic fields, negligible stray fields, and zero net topological charge, AFM skyrmions follow straight trajectories that prevent their annihilation at nanoscale racetrack edges. This makes the AFM skyrmions a more favorable candidate over the ferromagnetic (FM) skyrmion for future spintronic applications. This work proposes an AFM skyrmion-based neuron device exhibiting the leaky-integrate-fire (LIF) functionality by exploiting thermal gradient or alternatively perpendicular magnetic anisotropy (PMA) gradient in the nanotrack for leaky behavior by moving the skyrmion in the direction to minimize the system energy. Furthermore, it is shown that the AFM skyrmion couples efficiently to the soft ferromagnetic layer of a magnetic tunnel junction enabling efficient read-out of the skyrmion. The maximum change of 9.2% in tunnel magnetoresistance (TMR) is estimated for detecting the AFM skyrmion. Moreover, the proposed neuron device has the energy dissipation of 4.32 *fJ* per LIF operation thus, paving the path for developing energy-efficient devices in antiferromagnetic spintronics for neuromorphic computing.

## I. INTRODUCTION

The concept of magnetic skyrmions was proposed by Tony Skyrme in 1961 to elucidate the stability of hadrons in particle physics [1]. The emergence of spin textures in magnetic materials with non-centrosymmetry has been predicted theoretically since 1989 [2-5]. In 2009, skyrmions were first observed as a hexagonal lattice in a chiral magnet MnSi owing to the existence of bulk Dzyaloshinskii–Moriya interaction (DMI) [6, 7]. However, these skyrmions mainly existed at low temperatures. Later, the skyrmions were discovered in non-centrosymmetric B20 compounds such as MnSi [8], FeCoSi [9] and also in multi-layered structures (stacking of ferromagnet and heavy metal) with a unique topological charge $Q_c = 1/4\pi \int (\boldsymbol{m} \cdot \partial_x \boldsymbol{m} \times \partial_y \boldsymbol{m}) dxdy$ in the presence of interfacial DMI [7,10-12]. Here, *m* is the unit magnetization vector in the *xy* lattice space. Later, the skyrmions were found at the room temperature (RT) in magnetic multilayers like Ta/CoFeB/TaO$_x$ and Pt/CoFeB/MgO and were manipulated by various means including laser [12], thermal [13], external magnetic field [14], stress [15], voltage-controlled magnetic anisotropy [16], and spin-polarized electric current [17]. Among them, the thermal gradient acts as one of the promising ways of driving the magnetic skyrmion for energy-efficient spintronic devices [18].

Magnetic skyrmions have intriguing features such as inherent topological protection, nanoscale dimension (ranging from the 100 nm to the 1 nm) [19-21], high operating speed [21-23], and low-driving current density [24]. Because of these characteristics, skyrmions are the promising candidates for developing highly energy-efficient and ultra-dense integrated nanodevices. The current-induced skyrmion motion in the nanotrack has been widely explored to design skyrmion-based racetrack memory [25, 26], logic gates [27, 28], and transistors [29, 30] for von Neumann computing architecture. However, the rapid expansion of data volume has posed significant challenges to such computing systems in terms of their scalability and complexity [31]. To fix these issues, neuromorphic engineering acts as a viable and promising solution for the advancement of energy-efficient circuits and systems for specific tasks such as image and temporal pattern recognition. Such systems offer high processing speed and low-power consumption due to the parallel operation of artificial neurons and synapses that act as computing and storage elements, respectively [32-34]. This motivates the study of skyrmion-based devices for neuromorphic computing.

To date, the majority of studies have been focused on FM skyrmions based devices. However, due to skyrmion Hall effect (SkHE) [35] experienced by FM skyrmions, there is a displacement of such skyrmions in the in-plane direction, perpendicular to the direction it is driven along the track using spin-transfer torque/spin-orbit torque (SOT) mechanism. This pushes the skyrmion towards the nanotrack edge, causing it to annihilate. However, AFM skyrmions that are formed by two FM sublattices, coupled by the inter-sublattice exchange interaction with no net magnetic moment, completely eliminates the SkHE intrinsically. This allows efficient and ultrafast AFM skyrmion motion along the racetrack [36]. Moreover, AFM skyrmions are not susceptible to stray fields [37]. This makes AFM skyrmion a promising candidate for designing artificial neuron devices.

In the recent past, researchers developed a neuron model that accurately mimics a biological neuron known as LIF neuron model, a modified version of the original IF neuron [38]. Three distinct functionalities, namely integrating, leaking, and firing, are carried out by a LIF neuron. Throughout the integration phase, the input current spikes are given to the neuron that leads to an increase in the stored energy in the neuron. During the leaking phase, no input current spike is given to the neuron, thus gradually reducing the stored energy in the neuron over time. The neuron will finally release the total stored energy by firing an output spike once the accumulated energy hits a certain threshold level.

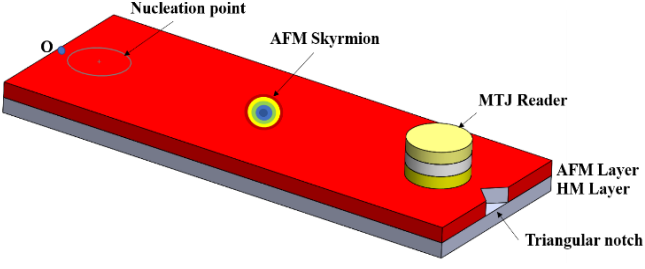

FIG. 1. Schematic diagram of the proposed device

Recently, it has been shown that artificial neurons based on phase-change memory [39], domain walls [40], and FM skyrmions [41] mimic the functionality of LIF neuron. Their behavior and performance, on the other hand, require some improvements. In 2017, Li *et al.*, suggested the FM skyrmion-based neuron device exhibiting the LIF functionality by employing the tunable current-driven skyrmion motion dynamics on a nanotrack having a PMA gradient [41]. Moreover, the proposed device was simulated at 0K (in the absence of thermal noise). However, for the practical implementation of skyrmion-based artificial neurons, a prerequisite is the stabilization of skyrmions at RT.

In this work, an AFM skyrmion-based artificial neuron device exhibiting the LIF functionality is proposed by exploiting the thermal gradient on a nanotrack as shown in Fig. 1. At first, the behavior and mechanism of AFM skyrmion dynamics on a nanotrack in the presence of thermal gradient is studied. Then, the performance of skyrmion dynamics on a nanotrack having the PMA gradient is also studied at non-zero temperature (RT). Further, it is shown that such mechanisms can be investigated in order to implement an LIF neuronal device with the distance moved by the AFM skyrmion along the nanotrack acting as biological neuron's membrane potential. The thermal/PMA gradient creates different energy states on the nanotrack relative to the skyrmion position. This causes an energy gradient on the nanotrack that further leads to a force acting on the AFM skyrmion, thus moving it backward to reduce the system energy and causing leaky behavior in the "LIF neuronal device." The suggested device's behavior and neural functionality are investigated using the micromagnetic model. The development of such a compact artificial neuron offers a novel way to achieve high-density and energy-efficient neuromorphic computing systems

## II.  METHODOLOGY

The proposed device is investigated using micromagnetic simulation tool "MuMax3" using the well-known Landau-Lifshitz-Gilbert (LLG) equation with spin-orbit-torque (SOT) as follows [42, 43]:

$$\frac{d\vec{m}}{dt} = -|\gamma|\vec{m} \times \vec{H}_{eff} + \alpha\left(\vec{m} \times \frac{d\vec{m}}{dt}\right) + \vec{\tau}_{SOT} \quad (1.1)$$

where $\vec{m}$ is the unit magnetization vector. $\alpha$ and $\gamma$ are the Gilbert damping coefficient and gyromagnetic ratio, respectively. $\vec{H}_{eff} = -\delta\vec{H}_{AFM}/(\mu_0\delta m)$ refers to the effective field that is associated with energies including DMI energy, PMA energy, exchange energy and thermal energy. Here, $\mu_0$ represents the absolute permeability of free space and $\vec{H}_{AFM}$ refers to the effective magnetic field for AFM system. The spin–orbit torque $\vec{\tau}_{SOT}$ is expressed as,

$$\vec{\tau}_{SOT} = -\frac{\gamma}{1+\alpha^2}\left(\left(\vec{m} \times \vec{H}_{SOT}\right) - \alpha\vec{H}_{SOT}\right) \quad (1.2)$$

where $\vec{H}_{SOT}$ is the effective field induced by spin polarization and can be defined as $\vec{H}_{SOT} = H_0(\vec{m} \times \vec{\sigma}_{SHE})$. Here, $H_0 = \frac{\mu_0 J_c \theta_{SHE}}{\gamma e t_0 M_s}$, $e$, $M_S$ and $t_0$ are the electron charge, saturation magnetization, and thickness of the film, respectively. $J_c$, $\theta_{SHE}$, and $\sigma_{SHE}$ are the charge current perpendicular to the plane, spin hall angle and electron spin polarization unit vector, respectively.

The effective magnetic field used to investigate the AFM system on the nanotrack has the following components [42, 44]:

$$\vec{H}_{AFM} = \vec{H}_{exch} + \vec{H}_{DMI} + \vec{H}_{anis} + \vec{H}_{therm} \quad (1.3)$$

$\vec{H}_{exch}$ is the effective field owing to Heisenberg exchange interaction that is represented as follows:

$$\vec{H}_{exch} = 2\frac{A_{ex}}{M_s}\sum_i \frac{\vec{m_i}-\vec{m}}{\Delta_i^2} \quad (1.4)$$

where $A_{ex}$ is the exchange stiffness. $\Delta_i$ indicates the size of the cell in the direction of neighbor $i$, where $i$ spans over the closest neighbors of the central cell of magnetization $\vec{m}$. $\vec{H}_{DMI}$ is the effective field owing to interfacial DMI that is defined as follows:

$$\vec{H}_{DMI} = \frac{2D}{M_s}\left(\frac{\partial m_z}{\partial x}, \frac{\partial m_z}{\partial y}, -\frac{\partial m_x}{\partial x} - \frac{\partial m_y}{\partial y}\right) \quad (1.5)$$

where $D$ represents the DMI constant. $x$, $y$, and $z$ components of unit magnetization vector $\vec{m}$ are $m_x$, $m_y$, and $m_z$, respectively. $\vec{H}_{anis}$ is the effective field induced by PMA that is defined as follows:

$$\vec{H}_{anis} = \frac{2K_u}{\mu_0 M_s}(\vec{u}.\vec{m})\vec{u} \quad (1.6)$$

where $K_u$ is the 1st order anisotropic constant and $\vec{u}$ refers to a unit vector in the direction of anisotropy. The fourth term on the right-hand side of equation (1.3) is the thermal field induced by thermal noise. This field can be expressed as $\vec{H}_{therm\_0} = \vec{\eta}\sqrt{\frac{2\alpha k_B T}{\mu_0 M_s \gamma V \Delta t}}$. Here, $\eta$ is defined as a random variable having the

Gaussian distribution with zero mean and unit variance and independent in each of the three Cartesian coordinates obtained at each time step. $k_B, T, V,$ and $\Delta t$ are the Boltzmann constant, temperature, volume of each simulation cell, and time step, respectively [44, 45].

Table 1 Material parameters of the proposed device for micromagnetic simulations [22,36]

| Parameters | Values |
|---|---|
| Saturation Magnetization $M_S$ | $376 \, kA/m$ |
| Gilbert Damping $\alpha$ | 0.1 |
| Exchange Stiffness $A_{ex}$ | $-8.5 \times 10^{-12} \, J/m$ |
| Spin polarization rate $P$ | 0.4 |
| Non-adiabatic STT $\beta$ | 0.5 |
| Interfacial DMI $D_{ind}$ | $1.2 \times 10^{-3} \, J/m^2$ |
| Anisotropic Constant $K_u$ | $1.1 \times 10^5 \, J/m^3$ |
| Elementary Charge $e$ | $1.6 \times 10^{-19} \, C$ |

The dimension of the nanotrack is considered as 324x120x2 nm³ for the simulations. The mesh size is chosen to be 2x2x2 nm³ that is less than the exchange length. The proposed device is simulated based on the parameters of KMnF₃ as shown in Table 1.

### III. STRUCTURE OF THE DEVICE

The suggested AFM skyrmion-based artificial neuron device is presented in Fig. 1. The main part of the device is a nanotrack that consists of a heavy metal (HM) layer with an AFM layer on the top of it. The AFM skyrmion is created at a point $X_0 = f(x,y) = (100,0) \, nm$ with respect to origin (O) on the nanotrack that has a fixed length of 324 nm. In order to stabilize the AFM skyrmion, the interfacial DMI and PMA are induced at the AFM-HM layer interface. In the proposed device, as shown in Fig. 2(a), the temperature difference between the two ends of the nanotrack is maintained in order to generate the linear thermal gradient on the AFM nanotrack, that is defined as follows: $T(x) = T_0 + \nabla T \cdot x$, where $T_0$ is the RT (300K) at one end, $\nabla T$ is the temperature gradient with respect to nanotrack length and its default value is 0.36 $Knm^{-1}$ and $x$ represents the distance with respect to the origin, which ranges from 0 to 324 nm. In Fig. 2(b), the PMA for AFM nanotrack increases linearly with the nanotrack length i.e., $K_u(x) = K_{u0} + \nabla K_u \cdot x$, where $K_{u0} = 1.1 \times 10^5 Jm^{-3}$, $\nabla K_u$ is the increasing rate of PMA with respect to track length and its default value is 268.5 $Jm^{-3} nm^{-1}$. A magnetic tunnel junction (MTJ) is located at $X_T = f(x,y) = (280,0) \, nm$ for the detection of AFM skyrmion. Thus, the skyrmion has to move 180 nm from its point of creation to be detected under the MTJ.

### IV. OPERATION OF THE DEVICE

The AFM skyrmion is initially created and then the current is injected into the HM layer that generates the spin current using spin Hall effect and drives the AFM skyrmion with SOT from the nucleation point to the detection point. The motion of the skyrmion to reach the detection point is determined by an evaluation of the overall forces that are acting on the skyrmion across the nanotrack. For the device shown in Fig. 1, the driving force is due to the applied SOT current and the backward force exerting on the AFM skyrmion is by the virtue of the existence of different energy states under the effect of thermal/PMA gradient on the nanotrack.

In the proposed device, the thermal gradient on the AFM nanotrack is generated by creating the temperature difference between both the extreme ends of the nanotrack, thus making the left end hotter as compared to the right-end. The left side experiences more entropy in the spin-dynamics of the electrons in comparison to the right side of the nanotrack (colder side), thus generating the magnon spin current and diffusing towards the colder region that is explained by various macroscopic thermodynamic theories namely, spin-dependent Peltier effect [46], spin Nernst effect [47], thermal Hall effect [48], and spin Seebeck effect [49]. The magnon current exerts a spin torque on the AFM skyrmion in accordance with the conservation of angular momentum. This makes the skyrmion move in the opposite direction of magnon current, towards the hotter side in order to reduce the system free energy.

In the proposed device, the skyrmion will move from colder to hotter region only if the input current density is sufficient to counteract the force due to thermal energy gradient on the nanotrack, otherwise, the AFM skyrmion moves backward. The MTJ will detect the skyrmion when it reaches the detection point $X_T$ and the neuron 'fires' an output spike and then, resets. The triangular notch-like structure can be employed at the nanotrack end as the skyrmion can easily be annihilated by this notch [50]. This is done in order to reset the device. This current-driven skyrmion motion is identical to the 'leaky-integrate-fire' mechanism of the LIF neuronal model. The LIF neuronal behavior attained by exploiting the thermal gradient on the nanotrack is also compared with one having the PMA gradient on the nanotrack at RT [41]. In this case, the force is due to anisotropic (PMA) energy gradient instead of thermal energy gradient, which is the keystone of leaky behavior in the proposed device.

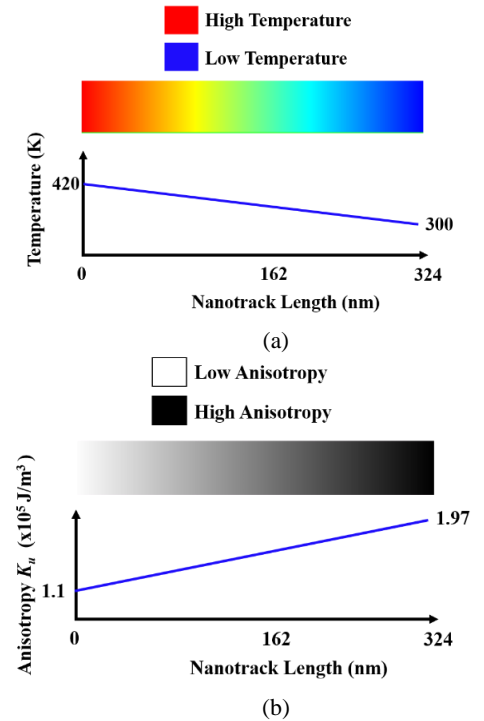

FIG. 2. Representation of a) thermal gradient b) PMA gradient on a nanotrack

## V. RESULTS AND DISCUSSION

The key idea behind the proposed AFM skyrmion based LIF neuron device is the tunability of skyrmion motion by employing the energy gradient on the nanotrack in the form of negative thermal gradient (decrease in temperature along the nanotrack length) and positive magnetic anisotropic gradient (increase in magnetic anisotropy along the nanotrack length). An AFM skyrmion is initially created at the point $X_0 = f(x,y) = (100,0) nm$ on the nanotrack. The SOT current drives the skyrmion along the nanotrack from the nucleation point to the detection point. During the skyrmion motion, there is a counteraction between the driving force produced by current and the force induced due to energy gradient existing on the nanotrack. The skyrmion moves forward whenever the driving force due to current overcomes the force induced by energy gradient; otherwise, it moves backward, thereby showing the "leaky" behavior of the LIF neuronal device. The neuron fires (produce an output signal) if the skyrmion reaches the MTJ, and is then reset using the triangular notch-like structure at the end of the nanotrack [50]. This behavior of skyrmion motion demonstrates the LIF functionality of the neuron device.

Fig. 3 (a) and 3 (b) shows the skyrmion motion in the presence of homogeneous and non-homogenous input current, respectively on a nanotrack having the thermal gradient.

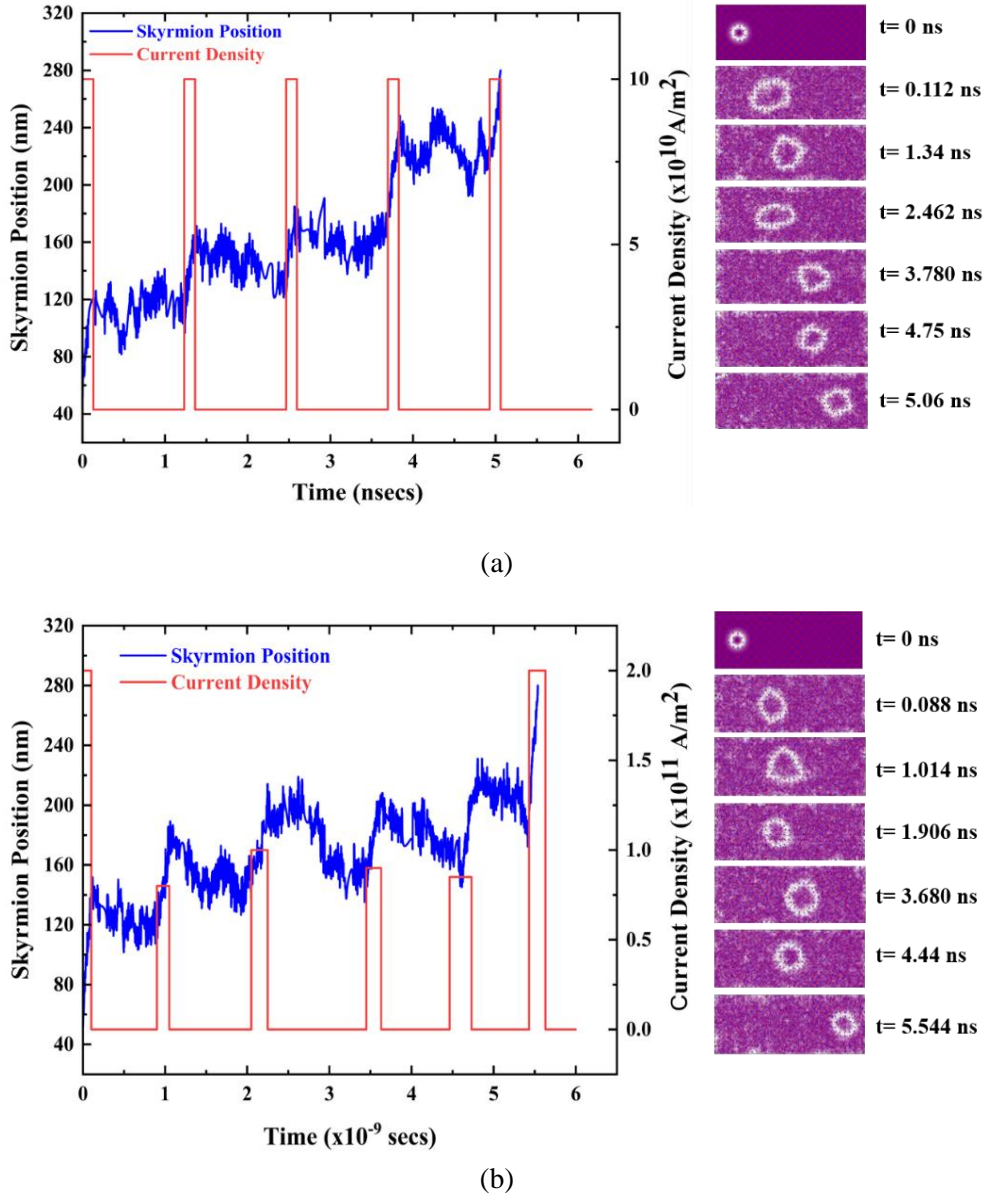

FIG. 3. Skyrmion position with respect to time having a thermal gradient on a nanotrack in presence of a) homogenous current pulse and b) inhomogeneous current pulse, showing LIF functionality.

In Fig. 3(a), the amplitude of uniform input current is considered as 1x10$^{11}$ A/m$^2$, frequency 0.81 GHz, and duty cycle of 10.71%. Fig. 4(a) and 4(b) depict the behavior of motion of the skyrmion in consideration with homogeneous and non-homogenous input current, respectively on a nanotrack having the PMA gradient. In Fig. 4(a), the amplitude of uniform input current is 2x10$^{11}$ A/m$^2$, frequency 1 GHz, and duty cycle of 20%. However, in Fig. 3(b) and 4(b), there is a random variation in amplitude and frequency of input current. The blue curve shows the distance moved by the skyrmion, exhibiting the LIF behavior with respect to time for both homogenous and non-homogenous current sequences. Fig. 3 and 4 also illustrates the micromagnetic simulations of the proposed device under different strategies (application of homogenous and non-homogenous current pulse) for thermal and PMA gradient, respectively on the nanotrack.

Considering the same set of parameters, a total of 25 micromagnetic simulations were carried out in the presence of thermal gradient $\nabla T = 0.36\, Knm^{-1}$ on the nanotrack as shown in Fig. 5. The stochastic realizations were evaluated for the variation in position of the skyrmion with respect to the time when the homogenous current pulse of 1x10$^{11}$ A/m$^2$, frequency of 0.81 GHz, and duty cycle of 10.71% is applied. The bold blue curve shows the mean of the stochastic simulations of position of the skyrmion with respect to time. In accordance with the thermal fluctuations, at short time scales, the skyrmion oscillates around an average position whereas in the long run, the skyrmion moves in the particular direction.

Fig. 5 shows the firing signal $f_{fire}$ of the proposed neuron device and can be expressed as follows:

$$f_{fire}(x,t) = K\delta(X_T - x(t,T)) = \begin{Bmatrix} K, x(t,T) = X_T \\ 0, \; x(t,T) \neq X_T \end{Bmatrix} \quad (1.7)$$

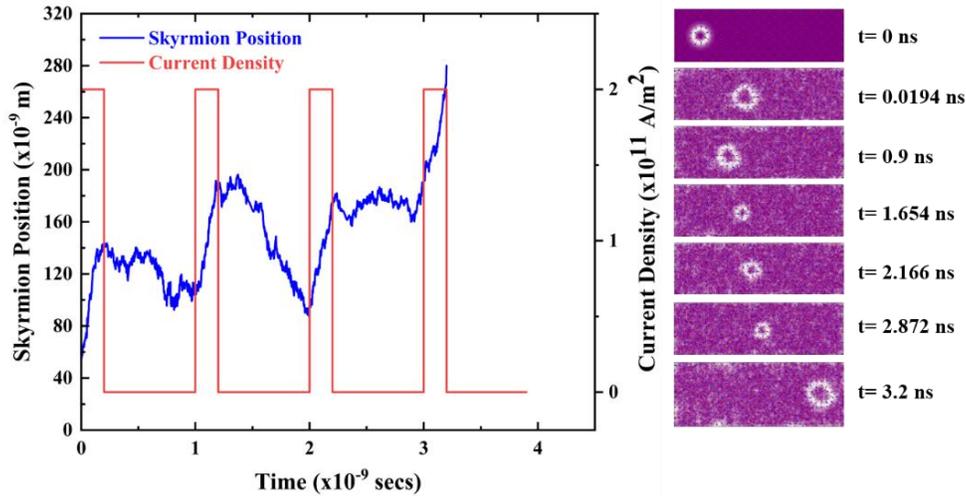

(a)

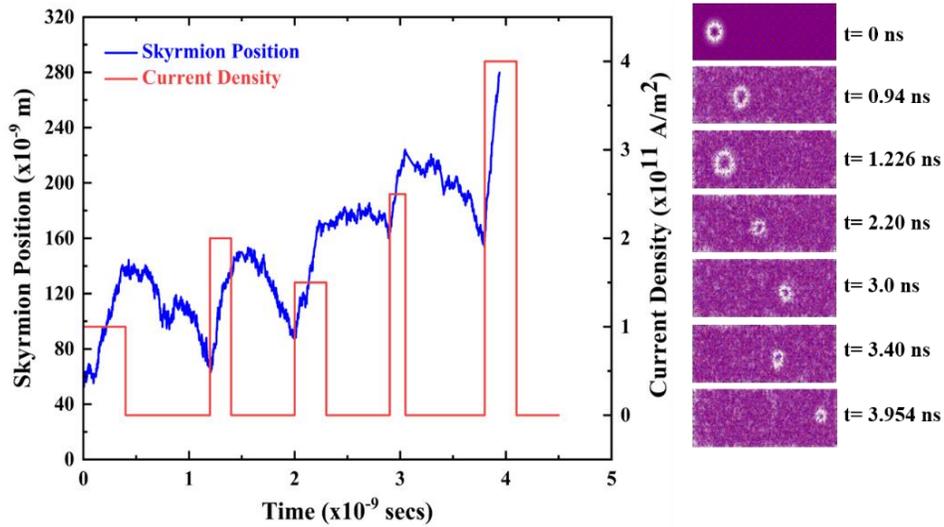

(b)

FIG. 4. Skyrmion position with respect to time having a PMA gradient on a nanotrack in presence of a) homogenous current pulse and b) inhomogeneous current pulse, showing LIF functionality.

where $K$ is the constant that represents the firing strength, $\delta$ refers to the Dirac-delta function and $x(t,T)$ is the distance moved by the skyrmion at any instant in the presence of thermal noise at temperature $T$. Whenever, the proposed neuron attains the maximum threshold position $X_T$, the "successful firing" will be indicated by the firing function. On the other hand, if the moving skyrmion doesn't reach the threshold distance $X_T$, then "no firing".

As shown in Fig. 1, the AFM skyrmion is detected using MTJ reader through TMR effect. The TMR is considered to follow an empirical $cos(\theta)$ law, where theta ($\theta$) is the angle between the magnetization in the free layer and fixed layer of MTJ [51]. In the presence of inter-layer exchange coupling between the free layer of MTJ (non-zero magnetic moment) and the AFM layer (zero net-magnetic moment), the local magnetization (overlapping region between the free layer of MTJ and AFM layer) results in the non-zero magnetic moment [52]. The change in spin angle of magnetization in AFM layer directly changes the relative magnetization between the fixed layer and free layer of MTJ, thereby leading to change in TMR in the presence/absence of AFM skyrmion.

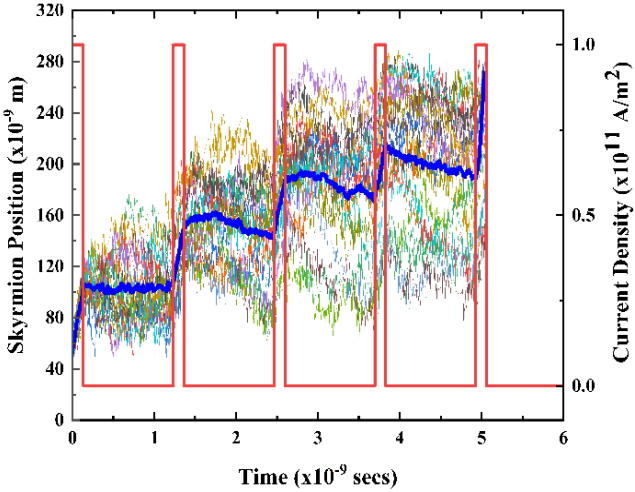

FIG. 5. Stochastic behavior of AFM skyrmion in the presence of thermal gradient on the nanotrack when homogenous current pulse is applied

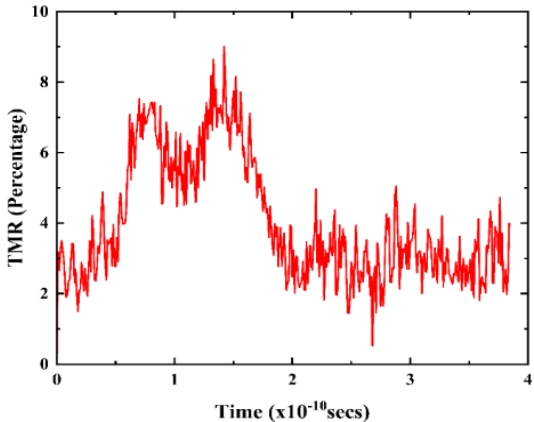

FIG. 6. Percentage change in TMR for the detection of AFM skyrmion

In Fig. 6, the change of TMR with respect to the time is shown. It is worthy to note that the two spikes are observed while detecting the AFM skyrmion. The first spike represents the entering of the AFM domain wall (present between the boundary and core of the AFM skyrmion) under the MTJ reader. The second spike is observed when the AFM domain wall that is present between the core and boundary of the AFM skyrmion starts exiting from the MTJ reader [53]. The TMR can be expressed as follows [54]:

$$TMR\ (\%age) = \frac{P^2(1-cos\theta)}{1-P^2 cos\theta} \times 100 \quad (1.8)$$

where $P$ refers to the spin polarization factor of free and fixed layer of the MTJ. Assuming $P$ of MTJ as 0.7, a 9.2% maximum change in TMR is estimated with respect to the parallel state of MTJ (the magnetization of fixed and free layer of MTJ is considered in +z direction). In this way, the detection of AFM skyrmion is achieved.

In addition, with an input spike current density of $10 \times 10^{10}$ Am$^{-2}$, average current density is $1.32 \times 10^{10}$ Am$^{-2}$. Assuming the width and thickness of HM to be 120 nm and 3nm, respectively, the average current flowing through the HM $I_{HM} = 4.75\mu A$. The process time ($t_p$) is 5 ns and the resistivity of the HM ($\rho_{HM}$) is assumed to be 1800 nΩm [55]. Consequently, the estimated total energy dissipation is 4.32 $fJ$, which is about 500 times lower than that (9.3pJ) of silicon-based spiking neuron [55]. This total energy includes the electrical energy (driving the skyrmion), thermal energy and read-out energy.

The electrical energy and read-out energy can be defined as: $E_e = I_{HM}^2 R_{HM} t_p = 1.82 fJ$ and $E_r = \int_0^{t_r} V_r I_r dt = 0.8 fJ$ respectively. Here, $V_r$ and $I_r$ are the read-circuit voltage and current, respectively. $t_r = 2ns$ is the read-out pulse width.

The thermal energy can be expressed as: $E_{th} = KA(\Delta T/\Delta x)\, t_p$, where $K = 40 mWatts °C^{-1} cm^{-1}$ is the thermal conductivity of KMnF$_3$ [56], $A = 120 \times 2\ nm^2$ is the cross-sectional area of AFM layer, $\Delta T = 120K$ is the temperature difference between the two ends, $\Delta x = 324 nm$. Out of the total energy, *1.7fJ* (thermal energy) is consumed in order to maintain the temperature gradient in the proposed device. Moreover, the thermal energy can be avoided if the waste heat is used to create the gradient on the nanotrack. These advantages of the proposed device could further minimize the energy consumption of emerging neuromorphic computing systems.

## VI. CONCLUSION

The stabilization of skyrmions at RT is a prerequisite for the practical implementation of skyrmion-based artificial neurons. The neuron device that exhibits LIF functionality has been proposed by exploiting the thermal gradient (or equivalently PMA gradient) on the nanotrack that acts as a promising and viable technique of driving the magnetic skyrmion. Under thermal gradient, the magnons that are the quanta of spin waves drive the skyrmion without Joule heating by virtue of absence of charge transport that makes them desirable for energy-efficient spintronics. In this work, the LIF neuronal behavior is demonstrated and its performance is analyzed through micromagnetic simulations. The thermal/PMA gradient creates different energy states on the nanotrack relative to the skyrmion position. This exerts a force on the skyrmion thus, driving it backwards so as to reduce the free energy, which is the basis of "LIF neuronal device." Moreover, the detection of AFM skyrmion is also realized through interlayer exchange coupling between the free layer of MTJ and the AFM layer and 9.2% of maximum change in TMR is observed. Furthermore, the energy dissipation of the proposed device is estimated as $4.32 fJ$ per LIF operation. Hence, the suggested energy-efficient artificial neuron highlights the potential applications of the AFM skyrmions in cutting-edge neuromorphic computing systems.

## ACKNOWLEDGEMENTS

N.B., R. K. R. and B. K. K would like to acknowledge Science and Engineering Research Board (SERB), Department of Science and Technology, Government of India under Grant CRG/2019/004551 for providing the funding to carry out the research work. M. M. R. and J. A. would like to acknowledge US National Science Foundation CISE SHF Small grant # 1909030.